\documentstyle[12pt,aaspp]{article}
\input{psfig.tex}
\begin{document}
\title{Microwave Background Constraints on Cosmological Parameters}
\normalsize
\author{Matias Zaldarriaga\footnote{matiasz@arcturus.mit.edu}}
\affil{Department of Physics, MIT, Cambridge, Massachusetts~~02139}
\author{David N. Spergel\footnote{dns@astro.princeton.edu}}
\affil{Department of Astrophysical Sciences, 
Princeton University, Princeton, New Jersey~~08544}
\author{Uro\v s Seljak\footnote{useljak@cfa.harvard.edu}}
\affil{Harvard-Smithsonian Center for Astrophysics,
60 Garden Street, Cambridge, Massachusetts~~02138}

\vspace{0.3 cm}

\begin{abstract}

We use a high-accuracy
computational code
to investigate the precision with which cosmological parameters
could be reconstructed by future cosmic microwave 
background experiments. We focus on the   
two planned satellite missions:
MAP and Planck.
We identify several 
parameter combinations that could be determined with a few percent 
accuracy with the two missions, as
well as some degeneracies among the parameters
that cannot be accurately resolved with the 
temperature data alone. These degeneracies can be broken
by other astronomical measurements. 
Polarization measurements can significantly enhance the
science return of both missions by allowing a more  accurate
determination of some cosmological parameters, by enabling
the detection of gravity waves and by probing the
ionization history of the universe.
We also address the question of how gaussian the likelihood
function is
around the maximum and whether 
gravitational lensing changes the constraints.

\end{abstract}

\newpage

\def\edth{\;\raise1.0pt\hbox{$'$}\hskip-6pt\partial\;}
\def\baredth{\;\overline{\raise1.0pt\hbox{$'$}\hskip-6pt
\partial}\;}
\def\bi#1{\hbox{\boldmath{$#1$}}}
\def\gsim{\raise2.90pt\hbox{$\scriptstyle
>$} \hspace{-6.4pt}
\lower.5pt\hbox{$\scriptscriptstyle
\sim$}\; }
\def\lsim{\raise2.90pt\hbox{$\scriptstyle
<$} \hspace{-6pt}\lower.5pt\hbox{$\scriptscriptstyle\sim$}\; }

\section{Introduction}

Measurements of 
cosmic microwave background (CMB) anisotropies have already 
revolutionized cosmology by providing insight
into the physical conditions of the universe only three
hundred thousand years 
after the Big Bang. The first year COBE data (\cite{smoot}) determined
the amplitude of the large angular scale CMB power spectrum  
with an accuracy of 10\% and the spectral slope with 
an accuracy of $0.3$. With the 4-year COBE data (\cite{bennett}),
these constraints became the tightest cosmological 
constraints available. 
Recent results from over a dozen balloon and ground based experiments 
are beginning to explore the anisotropies on smaller angular 
scales, which will help to constrain 
other cosmological parameters as well. The future looks even more
promising: there are now two planned satellite missions, 
MAP\footnote{See the MAP homepage at
http://map.gsfc.nasa.gov.}
and  
Planck\footnote{See Planck homepage at 
http://astro.estec.esa.nl/SA-general/Projects/Cobras/cobras.html}.
Both missions will provide a 
map of the whole sky with a fraction of a degree angular resolution
and sufficient signal to noise to reconstruct the underlying power 
spectrum with an unprecedented accuracy. 

A wonderful synergy is taking
place in the study of the cosmic background radiation:  theorists
are able to make very accurate predictions of CMB anisotropies
(Bond \& Efstathiou 1987; Hu et al. 1995; Bond 1996; 
Seljak \& Zaldarriaga 1996), while experimentalists
are rapidly improving our ability to measure these anisotropies.
If the results are consistent with structure
formation from adiabatic curvature fluctuations (as predicted
by inflation), then they can be used to accurately
determine a number of cosmological parameters.  
If the measurements are not consistent with any 
of the standard models, cosmologists will need
to rethink ideas about the origin of structure formation in 
the universe. In either
case, CMB will provide information about fundamental 
properties of our universe.

It has long been recognized that the 
microwave sky is sensitive to 
many cosmological parameters, so that a high resolution
map may lead to their accurate determination 
(\cite{bond94}; \cite{spergel}; \cite{knox}; \cite{jungman};
\cite{bondmax97}). 
The properties of the microwave background fluctuations are sensitive to 
the geometry of the universe, the baryon-to-photon ratio,
the matter-to-photon ratio, 
the Hubble constant, the cosmological constant, and the optical
depth due to reionization in the universe. 
A stochastic background of gravitational waves also leaves an imprint on 
the CMB and their amplitude and slope
may be extracted from the observations.
In addition,  
massive neutrinos and a change in the slope 
of the primordial spectrum also lead to potentially observable features.

Previous calculations trying to determine how well the various
parameters could be constrained were based on approximate
methods for computing the CMB spectra (\cite{jungman}). 
These approximations have an accuracy of several percent, which
suffices for the analysis of the present-day data. 
However, 
the precision of the future missions will be so high that the use  of such
approximations will not be sufficient for an accurate determination of
the parameters. 
Although high accuracy calculations are not needed at present to 
analyze the observations, they are needed to determine how 
accurately cosmological parameters can be extracted from a given 
experiment.
This is important not only for illustrative purposes but may also help
to guide the experimentalists in the design of the detectors. 
One may for example address the question of how much improvement one 
can expect by increasing the angular resolution of an experiment (and 
by doing so increasing the risk of systematic errors) 
to decide whether this 
improvement is worth the additional risk. Another question of current 
interest is whether it is worth sacrificing some sensitivity in the
temperature maps to gain additional information from the polarization of
the microwave background.
When addressing these questions,
the shortcomings of approximations become particularly problematic. 
The sensitivity to a certain parameter depends on  
the shape of the likelihood function around the maximum, which 
in the simplest approach used so far 
is calculated by 
differentiating the spectrum with respect
to the relevant parameter. This differentiation strongly amplifies  
any numerical inaccuracies: this
 almost always leads to an unphysical breaking of
degeneracies among  parameters and  misleadingly 
optimistic results. 

Previous analysis of CMB sensitivity to cosmological parameters used
only temperature information. However, CMB experiments can measure not
just the temperature fluctuations, but also even weaker variations in 
the polarization of the microwave
sky. Instead of one power spectrum,  
one can measure up to four and so increase 
the amount of information in the two-point correlators 
(\cite{uros,spinlong}, Kamionkowski, Kosowsky \& Stebbins 1997). 
Polarization can provide particularly 
useful information regarding the ionization history
of the universe (\cite{zal}) and the presence of a tensor contribution
(\cite{letter,kks}). Because these parameters 
are partially degenerate with others,
any improvement in their determination leads to a better reconstruction
of other parameters as well.
The two proposed satellite missions
are currently investigating the possibility
of adding or improving their ability to measure polarization, so it is
particularly interesting to address the question of improvement in the 
parameter estimation that results from polarization.

The purpose of this paper is to re-examine the determination of
cosmological parameters by CMB experiments 
in light of the issues raised above. It
is particularly timely to perform such an analysis now, when the 
satellite mission parameters are roughly defined. We use
the best current mission parameters in the calculations and
hope that our study provides a useful guide for mission
optimization. 
As in previous work (\cite{jungman}), 
we use the Fisher information matrix 
to answer the question of how accurately parameters
can be extracted from the CMB data. This approach 
requires a fast and accurate method for calculating the spectra and
we use the CMBFAST
package (\cite{sz96a}) with an accuracy of about 1\%.  
We test the Fisher information  method by 
performing a more general exploration of the shape of the 
likelihood function around its maximum and find that this 
method is sufficiently accurate for the present purpose. 

The outline of this paper is the following: in \S 2, we present the methods
used, reviewing the calculation of theoretical spectra 
and the statistical methods to address the question of sensitivity 
to cosmological parameters. In \S 3, we investigate the  
parameter sensitivity that could be obtained 
using temperature information only and in \S 4,
we repeat this analysis using both temperature and polarization information. 
In \S 5, we explore the accuracy of the Fisher method by performing a more 
general type of analysis and investigate the 
effects of prior information in the accuracy of the reconstruction.
We present our conclusions in \S 6.

\section{Methods}

In this section, we review the methods used  to calculate
the constraints on different cosmological parameters 
that could be obtained by the future CMB satellite experiments. 
We start by reviewing the statistics of CMB 
anisotropies in \S2.1, where we also 
present the equations that need 
to be solved to compute the theoretical prediction for the spectra
in the integral approach developed by Seljak \& Zaldarriaga (1996a). 
In  \S2.2,
we discuss the Fisher information matrix approach,
as well as the more general 
method of 
exploring the shape of the likelihood function around the minimum.

\subsection{Statistics of the microwave background}

The CMB radiation field is characterized by a $2\times 2$ intensity 
tensor $I_{ij}$. The Stokes parameters $Q$ and $U$ are defined as
$Q=(I_{11}-I_{22})/4$ and $U=I_{12}/2$, while the temperature 
anisotropy is
given by $T=(I_{11}+I_{22})/4$. In principle the fourth  
Stokes parameter $V$ that describes circular polarization would also 
be needed, but in standard cosmological models 
it can be ignored because it cannot
be generated through the process of Thomson scattering,  the
only relevant interaction process. 
While the temperature is
a scalar quantity that can naturally be expanded in spherical 
harmonics, $(Q\pm iU)$  
should be expanded using spin weight $\pm \ 2$ harmonics,  
(\cite{spinlong,goldberg67,gelfand63}, see also \cite{kks} for an 
alternative expansion)
\begin{eqnarray}
T(\hat{\bi{n}})&=&\sum_{lm} a_{T,lm} Y_{lm}(\hat{\bi{n}}) \nonumber \\
(Q+iU)(\hat{\bi{n}})&=&\sum_{lm} 
a_{2,lm}\;_2Y_{lm}(\hat{\bi{n}}) \nonumber \\
(Q-iU)(\hat{\bi{n}})&=&\sum_{lm}
a_{-2,lm}\;_{-2}Y_{lm}(\hat{\bi{n}}).
\label{Pexpansion}
\end{eqnarray}
The expansion coefficients 
satisfy $a_{-2,lm}^*=a_{2,l-m}$ and
$a_{T,lm}^*=a_{T,l-m}$.
Instead of $a_{2,lm}$ and  $a_{-2,lm}$ it is convenient to introduce their
even and odd parity linear combinations
(\cite{np})
\begin{eqnarray}
a_{E,lm}=-(a_{2,lm}+a_{-2,lm})/2 \nonumber \\ 
a_{B,lm}=i(a_{2,lm}-a_{-2,lm})/2.
\label{aeb}
\end{eqnarray}
These two combinations 
behave differently under parity transformation:  
while $E$ remains unchanged $B$ changes sign, in 
analogy with electric and magnetic fields.

The statistics of the CMB are characterized by the power spectra of
these variables together with their cross-correlations. 
Only the cross correlation between $E$ and $T$ is expected to be
non-zero, as $B$ has the opposite parity to both $E$ and $T$. 
The power spectra are defined as the rotationally invariant quantities
\begin{eqnarray}
C_{Tl}&=&{1\over 2l+1}\sum_m \langle a_{T,lm}^{*} a_{T,lm}\rangle 
\nonumber \\
C_{El}&=&{1\over 2l+1}\sum_m \langle a_{E,lm}^{*} a_{E,lm}\rangle 
\nonumber \\
C_{Bl}&=&{1\over 2l+1}\sum_m \langle a_{B,lm}^{*} a_{B,lm}\rangle 
\nonumber \\
C_{Cl}&=&{1\over 2l+1}\sum_m \langle a_{T,lm}^{*}a_{E,lm}\rangle. 
\label{Cls}
\end{eqnarray}

The four spectra above contain all the information on a given 
theoretical model, at least for the class of models described 
by  gaussian random fields. To test the sensitivity of the
microwave background to a certain cosmological parameter, 
we have to compute the  theoretical predictions for these spectra. 
This can be achieved by evolving the system of Einstein, fluid, and
Boltzmann
equations from an early epoch to
  the present. The solution for the spectra above
can be written as a line-of-sight integral over  sources 
generated by these initial perturbations
(\cite{sz96a}). For density perturbations (scalar modes),
the power spectra for $T$ and $E$  and their  cross-correlation 
are given by (the odd parity mode $B$ being zero in this case), 
\begin{eqnarray} 
C_{T,El}^{(S)}&=&
(4\pi)^2\int k^2dkP_\phi(k)\Big[\Delta^{(S)}_{T,El}(k)\Big]^2
\nonumber \\
C_{Cl}^{(S)}&=&
(4\pi)^2\int k^2dkP_\phi(k)\Delta^{(S)}_{Tl}(k)
\Delta^{(S)}_{El}(k),
\label{esc}
\end{eqnarray}
The source terms can be written as an integral over conformal time
(\cite{sz96a}, \cite{spinlong}),
\begin{eqnarray}
\Delta^{(S)}_{Tl}(k)&=&\int_0^{\tau_0} d\tau 
S^{(S)}_{T}(k,\tau) j_l(x) \nonumber \\  
\Delta^{(S)}_{El}(k)&=&\sqrt{(l+2)! \over (l-2)!}\int_0^{\tau_0} d\tau 
S^{(S)}_{E}(k,\tau) j_l(x) \nonumber \\  
S_T^{(S)}(k,\tau)&=&g\left(\Delta_{T,0}+2 \ddot{\alpha}
+{\dot{v_b} \over k}+{\Pi \over 4 }
+{3\ddot{\Pi}\over 4k^2 }\right)\nonumber \\
&+& e^{-\kappa}(\dot{\eta}+\ddot{\alpha})
+\dot{g}\left({v_b \over k}+{3\dot{\Pi}\over 4k^2 }\right)
+{3 \ddot{g}\Pi \over
4k^2} \nonumber \\
S^{(S)}_E(k\tau)&=& 
{3g(\tau)\Pi(\tau,k) \over 4 x^2}, \nonumber \\
\Pi&=&\Delta_{T2}^{(S)}
+\Delta_{P2}^{(S)}+
\Delta_{P0}^{(S)},
\label{es}
\end{eqnarray}
where $x=k (\tau_0 - \tau)$, $\tau_0$ is the present time 
and $\alpha=(\dot h + 6 \dot \eta)/2k^2$.
The derivatives are taken with respect to the conformal time $\tau$. 
The differential optical depth for Thomson scattering is denoted as 
$\dot{\kappa}=an_ex_e\sigma_T$, where $a(\tau)$ 
is the expansion factor normalized
to unity today, $n_e$ is the electron density, $x_e$ is the ionization 
fraction and $\sigma_T$ is the Thomson cross section. The total optical 
depth at time $\tau$ is obtained by integrating $\dot{\kappa}$,
$\kappa(\tau)=\int_\tau^{\tau_0}\dot{\kappa}(\tau) d\tau$.
We also introduced the visibility function $g(\tau)=\dot{\kappa}\ 
{\rm exp}(-\kappa)$. Its peak  
defines the epoch of recombination, which gives the  
dominant contribution to the CMB anisotropies.
The sources in these equations involve
the multipole moments of temperature and polarization, which 
are defined as $ \Delta(k,\mu)=\sum_l(2l+1)(-i)^{l}\Delta_l(k)P_l(\mu)$, 
where $P_l(\mu)$ is the Legendre polynomial of order $l$.
Temperature anisotropies have additional sources
in the metric perturbations $h$ and $\eta$
and in the baryon velocity term $v_b$ (Bond \& Efstathiou 1994; 1987; 
Ma \& Bertschinger 1996). The expressions above are only valid in 
the flat space limit; see Zaldarriaga, Seljak \& Bertschinger (1997)
for the appropriate generalization to open geometries.

The solution for gravity waves can be similarly written as
(\cite{spinlong})  
\begin{eqnarray}
\Delta_{Tl}^{(T)}&=&\sqrt{(l+2)! \over (l-2)!}\int_0^{\tau_0} 
d\tau S_T^{(T)}(k,\tau){j_l(x) \over x^2} \nonumber \\
\Delta_{E,Bl}^{(T)}&=&\int_0^{\tau_0} d\tau S_{E,B}^{(T)}(k,\tau)j_l(x) \nonumber \\
S_T^{(T)}(k,\tau) &=& -\dot he^{-\kappa}+g\Psi \nonumber \\
S_E^{(T)}(k,\tau)&=&-g\left(\Psi-{\ddot{\Psi}\over k^2}+{2\Psi \over x^2}
-{\dot{\Psi}\over kx}\right)+\dot{g}\left({2\dot{\Psi}\over k^2}+
{4 \Psi \over kx}\right)+2\ddot{g}{\Psi \over k^2}
 \nonumber \\
S_B^{(T)}(k,\tau)&=&g\left({4\Psi \over x}+{2\dot{\Psi}\over k}\right)+
2\dot{g} {\Psi \over k} \nonumber \\
&\Psi & \equiv  \Biggl\lbrack
{1\over10}\tilde{\Delta}_{T0}^{(T)}
+{1\over 7}
\tilde {\Delta}_{T2}^{(T)}+ {3\over70}
\tilde{\Delta}_{T4}^{(T)}
 -{3\over 5}\tilde{\Delta}_{P0}^{(T)}
+{6\over 7}\tilde{\Delta}_{P2}^{(T)}
-{3\over 70}
\tilde{\Delta}_{P4}^{(T)} \Biggr\rbrack,
\label{et}
\end{eqnarray}
where the power spectra are obtained by integrating
the contributions over all the wavevectors as in equation (\ref{esc}).
Note that in the case of tensor polarization perturbations
both $E$ and $B$ are present and can be measured separately. 
Inflationary models predict no vector component and we do not
include a vector component in our analysis.

\subsection{The Fisher information matrix}

The Fisher information matrix is a measure of the width and shape of the
likelihood function around its maximum. Its elements are
defined as expectation values of the
second derivative of a logarithm of the likelihood function with 
respect to the corresponding pair of parameters.
It can be used to estimate the accuracy
with  which the parameters in the cosmological model could be reconstructed
using the CMB data (\cite{jungman}, \cite{tegmark96}). 
If only temperature information is
given then for each $l$ a derivative of the temperature
spectrum $C_{Tl}$ 
with respect to the parameter under consideration is computed
and this information is then summed over all $l$ weighted  
by ${\rm Cov }^{-1}(\hat{C}_{Tl}^2)$. 
In the more general case implemented here, 
we have a vector of four derivatives and the
weighting is given by the inverse of the covariance matrix,
\begin{equation}
\alpha_{ij}=\sum_l \sum_{X,Y}{\partial C_{Xl} \over \partial s_i}
{\rm Cov}^{-1}(\hat{C}_{Xl},\hat{C}_{Yl})
{\partial C_{Yl} \over \partial s_j}.
\end{equation}
Here $\alpha_{ij}$ is the Fisher information 
matrix, ${\rm Cov}^{-1}$ is the inverse of the covariance matrix,
$s_i$ are the cosmological parameters one would like to 
estimate and $X,Y$ stands for $T,E,B,C$. For each $l$, one has to
invert the covariance matrix and sum over $X$ and $Y$. The derivatives
were calculated by finite differences and the step was usually
taken to be about $5 \%$ of the value of each parameter. We explored the
dependence of our results on this choice and found that the effect
is less than  $10 \%$. This indicates that
the likelihood surface is  approximately  gaussian. Further tests of this
assumption are discussed in \S 5.  

The full covariance matrix between the power spectra estimators  
was presented in \cite{uros}, \cite{spinlong} and \cite{kks}.
The diagonal terms are given by
\begin{eqnarray}
{\rm Cov }(\hat{C}_{Tl}^2)&=&{2\over (2l+1)f_{sky}}({C}_{Tl}+
w_T^{-1}B_l^{-2})^2
\nonumber \\
{\rm Cov }(\hat{C}_{El}^2)&=&{2\over (2l+1)f_{sky}}({C}_{El}+
w_P^{-1}B_l^{-2})^2
\nonumber \\
{\rm Cov }(\hat{C}_{Bl}^2)&=&{2\over (2l+1)f_{sky}}({C}_{Bl}+
w_P^{-1}B_l^{-2})^2
\nonumber \\
{\rm Cov }(\hat{C}_{Cl}^2)&=&{1\over (2l+1)f_{sky}}\left[{C}_{Cl}^2+
({C}_{Tl}+w_T^{-1}B_l^{-2})
({C}_{El}+w_P^{-1}B_l^{-2})\right],
\label{eqn7}
\end{eqnarray}
while the non-zero off diagonal terms are
\begin{eqnarray}
{\rm Cov }(\hat{C}_{Tl}\hat{C}_{El})&=&{2\over (2l+1)f_{sky}}{C}_{Cl}^2
\nonumber \\
{\rm Cov }(\hat{C}_{Tl}\hat{C}_{Cl})&=&{2\over (2l+1)f_{sky}}{C}_{Cl}
({C}_{Tl}+w_T^{-1}B_l^{-2})
\nonumber \\
{\rm Cov }(\hat{C}_{El}\hat{C}_{Cl})&=&{2\over (2l+1)f_{sky}}{C}_{Cl}
({C}_{El}+w_P^{-1}B_l^{-2}).
\label{eqn8}
\end{eqnarray}
We have defined $w_{(T,P)}^{-1}=4\pi \sigma_{(T,P)}^2 / N_{pix}$ where
$\sigma_T$ and $\sigma_P$ are
noise per pixel in the temperature 
and either $Q$ or $U$ polarization measurements (they are assumed
equal) and $N_{pix}$ is the number of pixels.
We will also assume that noise is uncorrelated between different pixels
and between different polarization components
$Q$ and $U$. This is only
the simplest possible choice and more complicated noise correlations
arise if all the components are obtained from a single set of
observations. If both temperature and polarization are
obtained from the same experiment by adding and differenciating
the two polarization states, then $\sigma_T^2=\sigma_P^2/2$ and
noise in temperature is uncorrelated with noise in
polarization components.
The window function $B_l^{-2}$ accounts
for the beam smearing and in the gaussian approximation is given by
$B_l^{-2}=e^{l^2 \sigma_b^2}$, with $\sigma_b$ measuring the width of
the beam.
We introduced $f_{sky}$ as the
fraction of the sky that can be 
used in the analysis. In this paper we assume $f_{sky}=0.8$.
It should be noted that equations (\ref{eqn7}) and (\ref{eqn8}) are
valid in the limit of uniform sky coverage. 

Both satellite missions will measure in several frequency
channels with different angular resolutions: 
we combine them using $w_{(T,P)}=\sum w_{(T,P)}^c$, where
subscript $c$ refers to 
each channel component. 
For MAP mission we adopt a noise level $w_T^{-1}=(0.11 \mu K)^2$
and $w_P^{-1}=(0.15 \mu K)^2$ for 
the combined noise of the three highest frequency channels 
with  
conservatively updated MAP beam sizes:
$0.53^\circ$, $0.35^\circ$ and $0.25^\circ$. 
These beam sizes are smaller than
those in the MAP proposal and represent improved estimates
of MAP's resolution. 
The most recent estimates
of MAP's beam sizes are even smaller\footnote{See the MAP homepage at
http://map.gsfc.nasa.gov.}:
$0.47^\circ$, $0.35^\circ$ and $0.21^\circ$
For Planck, we assume $w_T^{-1}=(0.011 \mu K)^2$ and 
$w_P^{-1}=(0.025 \mu K)^2$, and combine the 
$140$ GHz and $210$ GHz 
bolometer channels.  
For Planck's polarization sensitivity,
we assume a proposed design in which eight out of 
twelve receivers in each channel 
have polarizers (Efstathiou 1996). The angular 
resolution at these frequencies is $0.16^\circ$ and 
$0.12^\circ$ FWHM.  We
also  explore the possible science return from an enhanced
bolometer system that achieves polarization sensitivity of $w_P^{-1} =
(0.015 \mu K)^2$.

In our analysis, we are assuming that foregrounds can be subtracted
from the data to the required accuracy.
Previous studies of temperature anisotropies have shown that this 
is not an overly optimistic assumption at least on large 
angular scales (e.g., \cite{tegef96}). 
On smaller scales, point source removal as well as secondary processes
may make extracting the signal more problematic. This would mostly 
affect our results on Planck, which 
has enough angular resolution to measure features in the spectrum to 
$l \sim 3000$. For this reason, we compared the results by 
changing the maximum $l$ from $3000$ to $1500$. We find that they 
change by less than $30 \%$, so that the conclusions we find
should be quite robust.  
Foregrounds for
polarization have not been studied in detail yet. Given that there
are fewer foreground sources of polarization and that polarization 
fractions in CMB and foregrounds are comparable, we will make the
optimistic assumption that 
the foregrounds can be subtracted
from the polarization data with sufficient accuracy as well. 
However, as we will show, most of the additional information from 
polarization comes from very large angular scales, where the 
predicted signal is very small. Thus, one should
take our results on polarization as preliminary,  
until a careful analysis of foreground subtraction in polarization
shows at what level can polarization signal be extracted.

The inverse of the Fisher matrix, $\alpha^{-1}$,
is an estimate of the
covariance matrix between  parameters and $\sqrt{(\alpha^{-1})_{ii}}$
approximates 
the standard error in the estimate of the parameter $s_i$.
This is the lower limit because 
Cram\' er-Rao inequality guarantees that for an unbiased estimator 
the variance on i-th parameter has to be equal to or 
larger than $\sqrt{(\alpha^{-1})_{ii}}$.   
In addition to the diagonal elements of $\alpha^{-1}$, we will also
use $2\times 2$ submatrices of $\alpha^{-1}$ to analyze 
the covariance between various
pairs of parameters. {\it The Fisher matrix 
depends not only on the experiment
under consideration, but also on the assumed family of models 
and on the number of parameters that are being extracted from 
the data.}
To highlight this dependence
and 
to assess how the errors on the parameters depend on these choices 
we will vary their number and
consider
several different underlying models.

\subsection{Minimization}

The Fisher information matrix approach assumes
that the shape of the likelihood function around the maximum can be
approximated by a gaussian. In this section, we drop this assumption and
explore directly the shape of the likelihood function.
We use the
PORT optimization routines (\cite{gay90}) to explore one direction
in parameter space at a time by fixing one parameter 
to a given value 
and allowing the minimization routine to explore the
rest of parameter space to find the minimum of 
$\chi^2=\sum_l \sum_{X,Y}(C_{Xl}-C_{Xl}^*)
{\rm Cov}^{-1}(\hat{C}_{Xl}\hat{C}_{Yl})
(C_{Yl}-{C}_{Yl}^*)$, where $C_{Xl}^*$ denotes the underlying spectrum. 
The value of
$\chi^2$ as a function of this parameter can be compared directly
with the Fisher matrix prediction, $\Delta\chi^2=(s_i-s_i^*)^2/
(\alpha^{-1})_{ii}$, where $s_i^*$ is the value of the 
parameter in the 
underlying 
model. This comparison  tests not only 
the shape of the likelihood function around the maximum but also the
numerical inaccuracies resulting from differentiating the spectrum 
with respect to the relevant parameter. 
The minimization method is also useful for finding explicit
examples of degenerate models, models with 
different underlying parameters but almost indistinguishable spectra.

The additional advantage of the minimization approach is that one can 
easily impose 
various prior information on the data in the form of constraints
or inequalities. Some of these priors  
may reflect theoretical prejudice on the part of the person performing 
the analysis, while others are likely to
be less controversial: for example, the requirement  that 
matter density, baryon density and optical depth are all positive.
One might also
be interested in incorporating priors into the estimation 
to take other 
astrophysical information into account, e.g., the limits on the 
Hubble constant or $\Omega_m$ from the local measurements. 
Such additional information can 
help to break some of the degeneracies present in the CMB data, as
discussed in \S 3. 
Note that prior information on the parameters 
can also be incorporated into the Fisher matrix analysis, but 
in its simplest formulation
only in the form of gaussian constraints and not in the 
form of inequalities.  

The main disadvantage of this more general analysis is the
computational cost. At each step the minimization routine has to 
compute  derivatives with respect to all the parameters to find
the direction in parameter space towards the minimum.
If the initial model is sufficiently close to the minimum, then 
the code
typically requires 5-10 steps to find it and to sample the 
likelihood shape this has to be repeated for several values of the 
parameter in question (and also for several parameters). This computational
cost is significantly higher than in the Fisher matrix approach, where 
the derivatives with respect to each parameter 
need to be computed only once. 

\section{Constraints from temperature data}

In this section, we investigate how measurements of the CMB temperature 
anisotropies alone 
can constrain different cosmological parameters. 
The 
models studied here are approximately normalized to COBE, which sets
the level of signal to noise for a given experiment.

We will start with models in a 
six-dimensional parameter space 
$\bi {s}_6=(C_{2}^{(S)},h,\Omega_\Lambda,\Omega_b,\tau_{ri},n_s)$, 
where the parameters are respectively, 
the amplitude of the power spectrum for scalar perturbations at
$l=2$ in units of $\mu K^2$, the Hubble 
constant in units of 100 km/s/Mpc, the  
cosmological constant and baryon density in units of 
critical density, the reionization optical depth and the slope of 
primordial density spectrum.
In models with a non zero optical depth, we assume that the
universe is instantaneously and fully reionized, so that the ionization 
fraction is 0 before redshift of reionization $z_{ri}$ and 1 afterwards. We limit to this 
simple case because only the total optical
depth $\tau_{ri}$ can be usefully constrained 
without polarization information. We discuss the more general case
when we discuss polarization below.

The underlying model is standard CDM 
$\bi{s}_6=(796,0.5,0,0.05,0.05,1.0)$. 
Our base model has an optical depth 
of $0.05$, corresponding to the epoch of reionization at
$z_{ri}\approx 13$. Models include  gravity waves, fixing 
the tensor amplitude using the consistency 
relation predicted by inflation $T/S=-7n_T$ and assuming 
a relation between the scalar and tensor spectral slopes
$n_T=n_s-1$ for $n_s<1$ and $n_T=0$ otherwise, 
which is predicted by the simplest models of inflation. 
The results for MAP 
are summarized in table 1.   
It is important to keep in mind that the parameters are highly 
correlated. 
By investigating confidence contour plots in 
planes across the parameter space, one can identify combinations 
of parameters that can be more accurately determined. Previous analytical 
work (\cite{uros94,husugi}) showed  that the physics of
the acoustic oscillations is mainly determined by two parameters,
$\Omega_bh^2$ and $\Omega_mh^2$, where $\Omega_m$ is the 
density of matter in units of critical density. 
There is an
approximately  flat direction in the three dimensional space
of $h$, $\Omega_b$ and $\Omega_m$: for example,
one can change 
$\Omega_m$ and adjust $h$ and $\Omega_b$ to keep
$\Omega_bh^2$ and $\Omega_mh^2$ constant, which will not change the 
pattern of acoustic oscillations.
This degeneracy can be broken in two ways. 
On large
scales, the decay of the gravitational potential 
at late times in $\Omega_m \ne 1$ models 
(the so called late time integrated Sachs-Wolfe or ISW term) 
produces an additional component in the  microwave 
anisotropy power spectrum, which depends only on $\Omega_\Lambda$ 
(\cite{kofman}). Because the cosmic variance (finite 
number of independent multipole moments) is large for small $l$,
this effect cannot completely break the degeneracy. 
The second way is through the change in the angular 
size of the acoustic 
horizon at recombination, which shifts all the features in the 
spectrum by a multiplicative factor.
Around $\Omega_m=1$, this shift is a rather weak function of 
$\Omega_m$ and scales approximately as $\Omega_m^{-0.1}$, leading 
to almost no effect at low $l$, but is increasingly more important
towards higher $l$. MAP is  sensitive to multipole moments
up to $l \sim 800$, where this effect is small.
Consequently MAP's ability to determine the cosmological constant
will mostly come from large scales and thus will be
limited by  the large cosmic variance. 
Planck has a higher angular resolution and significantly 
lower noise, so it is 
sensitive to the change in the angular size of the horizon. 
Because of this Planck can break the parameter degeneracy 
and determine the cosmological constant to a high precision, as 
shown in table 2.

\begin{figure}[t]
\centerline{\psfig{figure=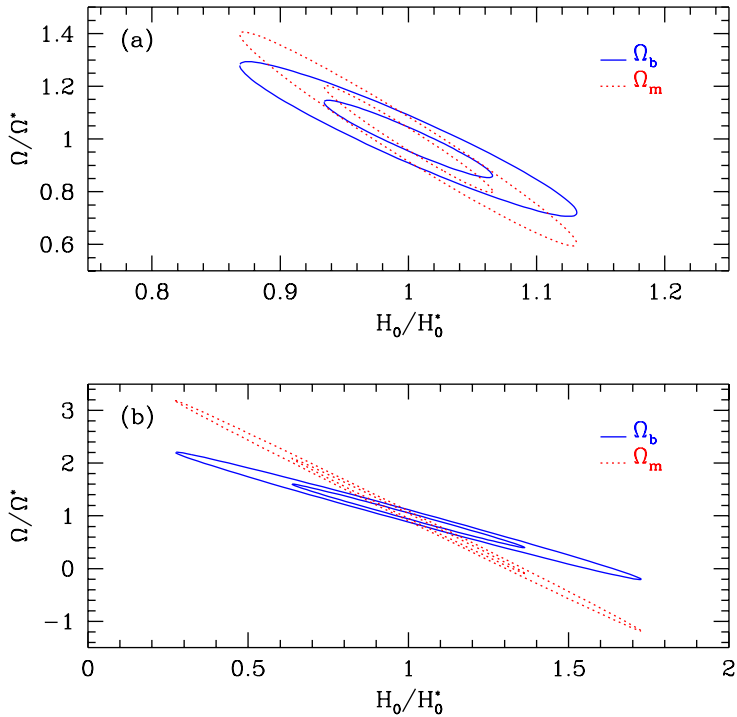,height=4in}}
\caption{MAP
confidence contours  
(68\% and 95\%)
for models in the six parameter space (a) and 
seven parameter space 
with $T/S$ added as a free parameter (b). Parameters are
normalized to their value in the underlying model denoted with
an asterisk.} 
\label{fig1}
\end{figure}

Figure \ref{fig1}a shows 
the confidence contours in the $\Omega_m - h$ and $\Omega_b - h$
planes. The error ellipses are significantly elongated along the lines
$\Delta\Omega_b/\Omega_b + 2.1 \Delta h /h=0$  and
$\Delta\Omega_m/\Omega_m + 3.0 \Delta h /h=0$. 
The combinations $\Omega_bh^{2.1}$ and $\Omega_mh^3$ 
are thus better determined than the parameters 
$\Omega_m$, $\Omega_b$ and $h$ 
themselves, both to about 3\% for MAP. 
It is interesting to note that it is $\Omega_mh^3$ rather 
than $\Omega_mh^2$ that is most accurately determined, which 
reflects the fact that ISW tends to break the 
degeneracy discussed above. However, because the ISW
effect  itself can be mimicked by 
a tilt in the spectral index the degeneracy remains, 
but is shifted to 
a different combination of parameters. One sigma standard errors
on the two physically motivated parameters 
are $\Delta (\Omega_bh^2) /\Omega_bh^2\approx 3 \%$ 
and $\Delta(\Omega_mh^2)/\Omega_mh^2\approx 5 \%$.
The fact that there is a certain degree of degeneracy between
the parameters has already been noted in previous work (e.g.,  
\cite{bond94}).

\begin{figure}[t]
\vspace*{6cm}
\caption{Confidence contours $(68\%\  \& \ 95\%)$ in the
$C_2^{(S)}-\tau_{ri}$ plane
for models in the six parameter space described in the text
with (dotted lines) 
or without (solid lines) polarization information. } 
\includegraphics{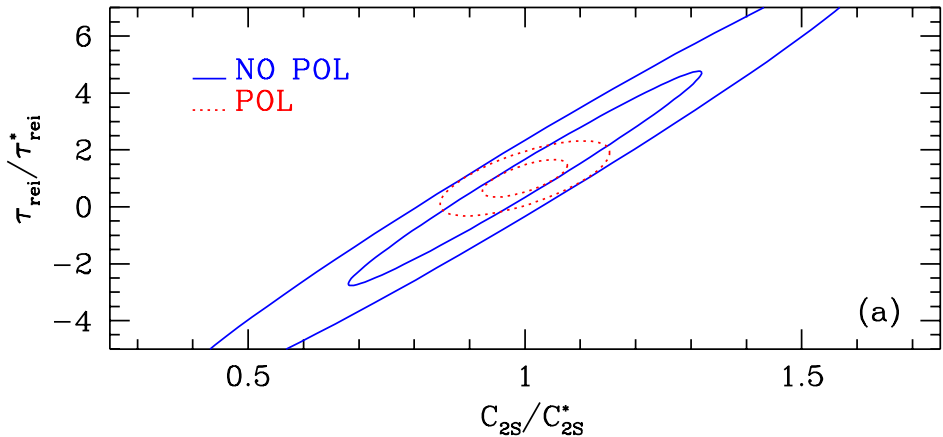}

\label{fig3}
\end{figure}

Another approximate 
degeneracy present in the temperature spectra is between 
the reionization optical depth $\tau_{ri}$ and amplitude $C_2^{(S)}$. 
Reionization uniformly suppresses the  
anisotropies from 
recombination by $e^{-\tau_{ri}}$. On large angular scales, new
anisotropies are generated during reionization 
by the modes that have not yet entered the horizon. 
The new anisotropies compensate the $e^{-\tau_{ri}}$ suppression, 
so that there is no suppression of anisotropies on COBE scales.
On small scales, the
modes that have entered the horizon have wavelengths 
small compared to the width of the new visibility function and so 
are suppressed because of cancellations between positive and
negative contributions along the line of sight and become
negligible.
The net result is that on small scales the spectrum is suppressed 
by $e^{-2\tau_{ri}}$ compared to the large scales.
To break the
degeneracy between $C_2^{(S)}$ 
and $\tau_{ri}$ one has to be 
able to measure the amplitude of the anisotropies 
on both large and small scales and
this is again limited on large scales by cosmic variance.
Hence one cannot accurately determine the two parameters separately, while
their combination $C_2e^{-2\tau_{ri}}$ 
is much better constrained. Figure \ref{fig3} shows that indeed the
error ellipsoid is very elongated in the direction $\Delta C_2/C_2
-0.1\Delta \tau_{ri}/\tau_{ri}=0$, which corresponds to 
the above combination  for $\tau_{ri}=0.05$.

We now allow for one more free parameter, the ratio of the tensor
to scalar quadrupole anisotropy
$T/S$, fixing the tensor spectral index $n_T$ using the consistency 
relation predicted by inflation $T/S=-7n_T$ but not assuming a relation
between $n_T$ and $n_s$. 
The variances 
for  MAP are again summarized in table 1. 
A comparison with the previous case shows that most variances
have increased.
Variances
for $h$ and  $\Omega_\Lambda$  are approximately five times
larger than before 
while that for $n_s$ has increased by a factor of six and that for
$\Omega_bh^2$ by almost four. On the other
hand, the error bar for $\tau_{ri}$ remains unchanged.
It is instructive to look again at the contour plots in the 
$\Omega_m - h$ and $\Omega_b - h$ planes shown in
Figure \ref{fig1}b. The degeneracy on individual parameters
is significantly worse because the large angular scale amplitude 
can now be adjusted freely with the new extra degree
of freedom, the tensor to scalar ratio $T/S$. This can therefore
compensate any large scale ISW term and so the degeneracy between 
$h$, $\Omega_\Lambda$ and $\Omega_b$  cannot be broken as easily.
However, a combination of the two parameters is still 
well constrained as shown in figure \ref{fig1}. 
The degenerate lines are now given by
$\Delta\Omega_b/\Omega_b + 1.66 \Delta h /h=0$  and
$\Delta\Omega_m/\Omega_m + 3.0 \Delta h /h=0$, with 
relative errors
$\Delta (\Omega_bh^{1.66}) /\Omega_bh^{1.66}\approx 4 \%$ 
and $\Delta(\Omega_mh^3)/\Omega_mh^3\approx 4 \%$, almost unchanged 
from the 6-parameter case. 
On the other hand for the physically relevant parameters
$\Omega_bh^2$ and $\Omega_mh^2$
we now have 
$\Delta (\Omega_bh^2) /\Omega_bh^2\approx 10 \%$ 
and $\Delta(\Omega_mh^2)/\Omega_mh^2\approx 25 \%$,
which is worse than before. This example indicates 
how the errors on individual parameters can change dramatically 
as we add more parameters while certain combinations of them 
remain almost unaffected.

\begin{figure}[t]
\centerline{\psfig{figure=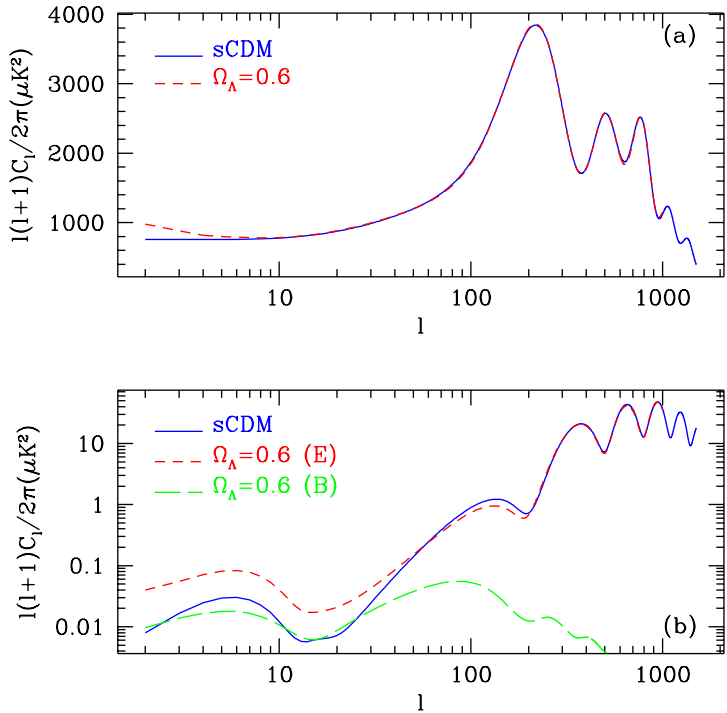,height=4in}}
\caption{Power spectra of (a) temperature and (b) polarization 
for two  
models that will be degenerate for MAP 
if only temperature information is used. 
The model with $\Omega_\Lambda=0.6$ is
the result of the minimization relative to the
sCDM for models constrained to have $\Omega_\Lambda=0.6$.
Polarization helps to break this degeneracy.} 
\label{fig2}
\end{figure}

The output of a minimization run trying to fit sCDM temperature
power spectra with models constrained to have 
$\Omega_{\Lambda}=0.6$ shows how different parameters 
can be adjusted in order to keep the power spectrum nearly the same.
The minimization program found the
model $\bi{s}_7=(610,0.67,0.6,0.03,0.09,1.1,0.68)$
where the last number now corresponds to the $T/S$ ratio, as
a model almost indistinguishable from the underlying one. 
The two models differ by $\Delta\chi^2=1.8$ and are
shown in figure \ref{fig2}.
It is interesting to analyze how  different parameters are adjusted
to reproduce the underlying model.
By adding gravity waves 
and increasing both the spectral index and the 
optical depth, the    
ISW effect from the cosmological constant can be compensated so that 
it is only noticeable for the first couple of $C_l$'s. 
The relatively high amount of tensors ($T/S \sim 0.7$)
lowers the scalar normalization
and thus the height
of the Doppler peaks, which is compensated by the increase in the
spectral 
index to $n_s=1.1$
and the decrease of $\Omega_m h^2$ from $0.25$ to 
$0.18$. The latter moves 
the matter radiation equality closer to recombination
increasing the height of the peaks. 
This is the reason why the degeneracy line is not that of 
constant
$\Omega_m h^2$ as figure \ref{fig1} shows. Changes in
$\Omega_m h^2$ change the structure of the peaks
and this can 
be compensated by changing other parameters 
like the optical depth 
or the slope of the primordial
spectrum. 
This cannot be achieved across all the spectrum
so one can expect that the degeneracy  
will be lifted as one increases the angular resolution,
which is what happens if Planck specifications are used 
(table 1).
Note that the amount of gravity waves
introduced to find the best fit does not follow the relation 
between $n_T$ and $n_s$ 
predicted by the simplest inflationary models 
discussed previously:
for $n_s=1.1$ no gravity waves are predicted.  
This explains why the addition of $T/S$ as a free parameter
increases the sizes of most error bars compared to the 6-parameter case. 

\begin{figure}[t]
\centerline{\psfig{figure=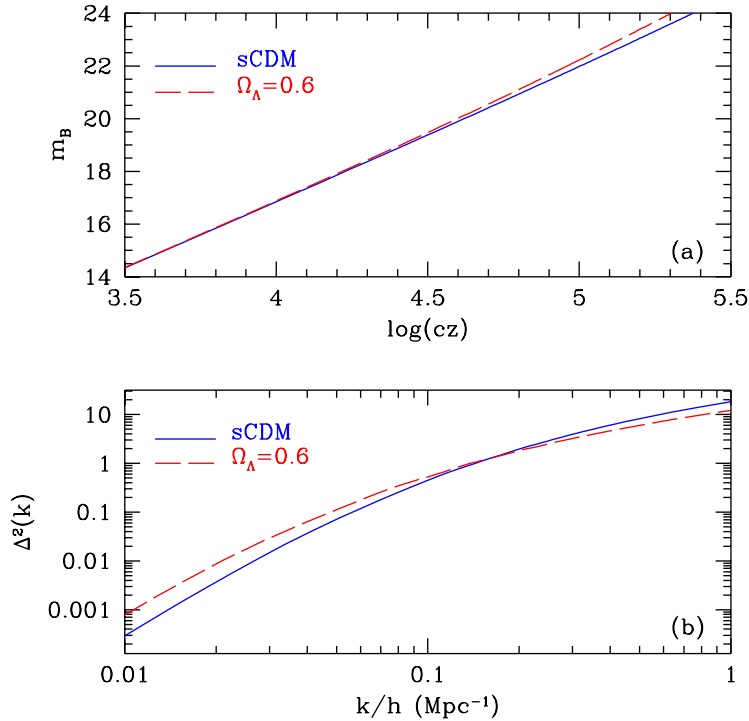,height=4in}}
\caption{Hubble diagram for Type Ia supernovae (a) and 
CDM linear power spectra (b) for sCDM and the $\Omega_\Lambda=0.6$ model
described in the text.} 
\label{fig4}
\end{figure}

While the two models shown in figure \ref{fig2} have very similar
temperature anisotropy spectra, they make very different 
astronomical predictions.  Figure \ref{fig4} shows the matter
power spectra of the two models.  An interesting effect is
that the two models are nearly identical on the scale
of $k = 0.1 h $ Mpc$^{-1}$, which corresponds to
$l \sim k\tau_0 \sim 600$, the $l$ range where MAP is very
sensitive and  gravity waves are unimportant.  However, the two
models differ significantly on the $0.01 h $ Mpc$^{-1}$
scale and the effective power spectrum 
shape parameter $\Gamma$ (e.g. Bardeen et al. 1986) 
is very different for the two models: 0.42 for the matter
dominated model and 0.25 for the vacuum dominated model.
The current observational situation is still controversial 
(e.g., \cite{peacock}), but
measurements of
the spectrum by the Sloan Digital Sky Survey (SDSS)
should significantly improve
the power spectrum determination.
The models also
make different predictions for cluster abundances:
the matter dominated model has $\sigma_8 \Omega_m^{0.6} = 1.2$,
while the vacuum dominated model has $\sigma_8 \Omega_m^{0.6}
= 0.8$. Analysis of cluster X-ray temperature and luminosity
functions suggests 
$\sigma_8  \Omega_m^{0.6}
= 0.5 \pm 0.1$ (\cite{eke}), inconsistent with both
of the models in the figure. 
These kind of measurements can break some of the 
degeneracies in the CMB data.

Observations of 
Type Ia supernovae at redshifts $z\sim 0.3-0.6$ is another
very promising way of measuring cosmological parameters. 
This test complements the CMB constraints
because the combination of $\Omega_m$ and $\Omega_\Lambda$ that leaves
the luminosity distance to a redshift $z\sim 0.3-0.6$ unchanged
differs from the one that leaves the position of the Doppler peaks
unchanged. Roughly speaking, the SN observations are sensitive to
$q_0 \simeq \Omega_m/2 - \Omega_\Lambda$, while the CMB observations are
sensitive to the luminosity distance which depends on a roughly
orthogonal combination, $\Omega_m + \Omega_\Lambda$.
Figure \ref{fig4}a 
shows the  apparent magnitude vs. redshift plot for
supernovae in the two models of figure 3.
The analysis 
in \cite{perlmutter} of the first seven supernovae already excludes the
$\Omega_\Lambda=0.6$ model with a high confidence.  However,
tt remains to be seen however whether this test will be free 
of systematics such as 
evolutionary effects
that have plagued other classical cosmological tests
based on the luminosity-redshift relation.

Finally, we may also relax the relation between tensor spectral index
and its amplitude, thereby testing the consistency relation of inflation.
For MAP, we studied two sCDM models, one
with $T/S=0.28$ and one with $T/S=0.1$, but with
$\tau_{ri}=0.1$, 
for Planck we only used the latter model.
Table 1 summarizes
the obtained one sigma limits.
A comparison between the $T/S=0.28$ model and previous results
for sCDM with seven parameters shows that the addition of $n_T$ as a new
parameter does not significantly change the expected sensitivities to
most parameters.
The largest change, as expected, 
is for the tensor to scalar ratio. We now find
$\Delta T/S\sim 0.7$ which means that the consistency 
relation will only be poorly  tested from the temperature 
measurements.  
If $T/S=0.1, \tau_{ri}=0.1$ most error bars
are smaller than if $T/S=0.28, \tau_{ri}=0.05$ case. The
reason for this is that the higher 
value of the optical depth in the underlying model makes its detection
easier and this translates to smaller error bars on the other parameters. 
The only exception is $n_T$, which has significantly higher error
if $T/S=0.1$ 
than if $T/S=0.28$ as expected on the basis  
of the smaller contribution of tensor modes to the total anisotropies.
A comparison between the expected MAP and Planck performances
for the $T/S=0.1$ model shows that Planck error bars 
are significantly smaller.
For $h$, $\Omega_bh^2$ and $\Omega_\Lambda$ the improvement is by a factor
of $10-20$, while for $T/S$ and $C_2^{(S)}$ by 
a factor of $2-3$. The limits on $\tau_{ri}$ and $n_T$ remain 
nearly unchanged, reflecting the fact that these parameters are 
mostly constrained on large angular scales which are cosmic variance
and not noise/resolution limited. It is for these parameters that 
polarization information helps significantly, as discussed in the 
next section.

The accuracy with which certain parameters can be determined 
depends not only on the number of parameters but
also on their ``true" value. We tested
the sensitivity of the results by repeating the analysis around a cosmological 
constant model $\bi{s}_8=(922,0.65,0.7,0.06,0.1,1.0,0.1,0.0)$, where 
the last number corresponds to the tensor spectral index $n_T$.
Results for MAP specifications are given in table 1.  
The most dramatic change is for the cosmological constant, which 
is a factor of ten better constrained in this case. This is 
because the underlying model has a large ISW effect which increases
the anisotropies at small $l$. 
This cannot be mimicked by adjusting the tensors,
optical depth and scalar slope as can be done if
the slope of the underlying model is flat, such as for sCDM model
in figure \ref{fig2}.
Because of the degeneracy between $\Omega_\Lambda$ and $h$ ,
a better constraint on the former will also improve the latter, as
shown in table 1. Similarly because a change 
in $\Omega_\Lambda$ affects $T/S$, 
$\tau_{ri}$ and $n_s$ on large scales, 
the limits on these parameters will
also change. 
On the other hand, errors on $\Omega_bh^2$, $C_2^{(S)}$ and $n_T$ 
do not significantly change. 
This example clearly shows that the effects of the 
underlying model can be rather significant for certain parameters, so 
one has to be careful in quoting the numbers without specifying the 
``true" parameters of the underlying model as well.

So far we only discussed flat cosmological models. CMBFAST can 
compute open cosmological models as well,
and we will now address the question 
of how well can curvature be determined using temperature data. 
We consider models in a six parameter space 
$\bi {s}_6=(C_{2}^{(S)},h,\Omega,\Omega_b, \tau_{ri},n_s)$, 
with no gravity waves and where
$\Omega_\Lambda=0$, so that $\Omega=\Omega_m$.
We will consider as the underlying
model $\bi{s}_6=(1122,0.65,0.4,0.06,0.05,1.0)$. 
Fisher matrix results are displayed in table 1. Within this family
of models $\Omega$ can be determined very precisely by both MAP
and Planck due to its effect on the position of the Doppler peaks.
This conclusion changes drastically if we also allow cosmological 
constant, in which case $\Omega=\Omega_m+\Omega_\Lambda$.
Both $\Omega$ and $\Omega_\Lambda$ change 
the angular size of the sound horizon at recombination 
so it is possible to change 
the two parameters without changing  
the angular size, hence the two parameters will be nearly
degenerate in general. 
We will discuss this degeneracy in greater detail in the next section, 
but we can already say  that including both parameters in the analysis
increases the error bar on the curvature dramatically.

To summarize our results so far, keeping in mind that 
the precise numbers depend on the underlying 
model and the number of parameters being extracted, we may 
reasonably expect that using temperature information only
MAP (Planck) will be able to achieve accuracies of
$\Delta C_2^{(S)}/C_2^{(S)} \sim 0.5 (0.1)$, 
$\Delta h \sim 0.1 (0.006)$, $\Delta \Omega_{\Lambda} \sim 0.6 (0.03)$,
$\Delta (\Omega_bh^2)/\Omega_bh^2 \sim 0.1(0.008) $,
$\Delta\tau_{ri} \sim 0.1 (0.1)$,
$\Delta n_s\sim 0.07 (0.006)$, $\Delta (T/S) \sim 0.7 (0.3)$ 
and $\Delta n_T \sim 1 (1)$. 
It is also worth 
emphasizing that there are combinations of the parameters that are 
very well constrained, 
e.g., $\Delta(\Omega_mh^3)/\Omega_mh^3\sim 0.04$ and 
$\Delta(C_2^{(S)})/C_2^{(S)}- 2\Delta{\tau_{ri}}\sim 0.05$. 
For the family of models with curvature but no cosmological 
constant, MAP (Planck) will be able to achieve 
$\Delta \Omega \sim 0.007 (0.0006)$, determining the curvature of 
the universe with an impressive accuracy.
 
These results agree qualitatively, but not 
quantitatively, with those in \cite{jungman}. The discrepancy
is most significant for 
$C_2^{(S)}$, $h$ and $\Omega_\Lambda$, for which 
the error bars obtained here
are significantly larger.  
The limit we obtain for  $\Omega_bh^2$ is 
several times smaller than that in \cite{jungman},  
while for the rest of the
parameters the results agree. The use of different codes
for computing model predictions is probably the main cause of 
discrepancies and emphasizes
the need to use high 
accuracy computational codes when performing 
this type of analysis. 

\section{Constraints from temperature and polarization data}

In this section, we consider the constraints on cosmological parameters
that could be obtained
when both temperature and polarization data are used.
To generate
polarization, two conditions have  to be satisfied: photons need to 
scatter (Thomson scattering has a polarization dependent
scattering cross-section) and the angular distribution of
the photon temperature must have a non-zero quadrupole moment. 
Tight
coupling between photons and electrons prior to recombination makes
the photon temperature distribution nearly isotropic and the
generated polarization very small, specially on scales larger than 
the width of the last scattering surface. For this reason
polarization 
has not been considered previously as being important for the determination 
of cosmological parameters. However,
early reionization increases the polarization amplitude
on large angular scales  in a way which 
cannot be mimicked with variations in other parameters
(\cite{zal}).
The reason for this is that after recombination 
the quadrupole moment 
starts to grow due to the photon free streaming. If there
is an early reionization with sufficient optical depth, then 
the new scatterings can transform this
angular anisotropy into polarization. This effect 
 dominates on the angular scale of  the horizon when reionization occurs.
It will produce a peak in the polarization power
spectrum with an amplitude proportional to the optical depth,
$\tau_{ri}$,
and a position,
$l\sim 2\sqrt{z_{ri}}$, where $z_{ri}$ is the redshift of
reionization.

We first consider
the six parameter space described in the previous section.
Table 2 
contains the one sigma errors on the parameters for MAP 
specifications. 
Compared to the temperature case,
the errors improve particularly 
on the amplitude, the reionization optical depth  
and the spectral index $n_s$. 
Figure \ref{fig3} shows the 
confidence contours for $C_2^{(S)}$ and $\tau_{ri}$ with and without 
polarization. One can see from this figure how
the information in the polarization breaks the degeneracy between the
two parameters by reducing the error on $\tau_{ri}$, 
but does not really improve their non-degenerate
combination, which is well determined from the temperature data alone.

We now allow for one more parameter,
$T/S$. Again, polarization improves the errors 
on most of the parameters 
by a factor of two compared to the no-polarization case,
as summarized in table 2.
The optical depth and the amplitude are better constrained for the same 
reason as for the six parameter model discussed above. 
Without polarization,
the extra freedom allowed by the gravity waves made it
possible to compensate the changes on large angular 
scales caused by the ISW, 
while the amplitude of small scale fluctuations could be adjusted
by changing the optical depth and the spectral index.
Changing $n_s$, also changes the slope on 
large angular scales 
compensating the change caused by the ISW. 
When polarization is included, a
change in the optical depth produces a large effect in the spectrum:
see the model  with $\Omega_\Lambda=0.6$ 
in figure \ref{fig3}, which has $\tau_{ri}=0.1$. 
The difference in $\chi^2$ between the two models in figure 
\ref{fig3} becomes 10 instead of 1.8. 

\begin{figure}[t]
\centerline{\psfig{figure=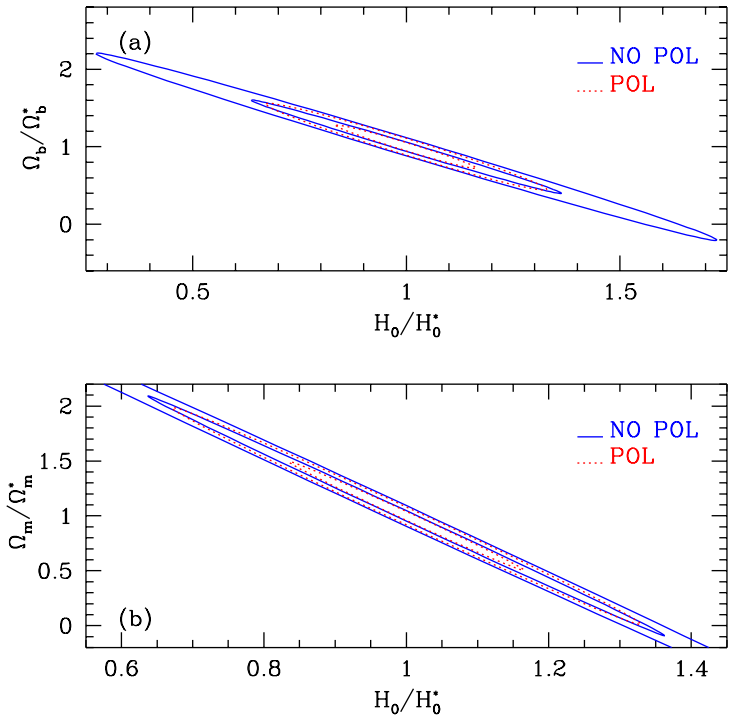,height=4in}}
\caption{Confidence contours $(68\%\  \& \ 95\%)$ in the
(a) $\Omega_b-h$ plane and (b) $\Omega_m-h$ plane
for models in the seven parameter space described in the text
with or without using polarization information. } 
\label{fig5}
\end{figure}

Figure \ref{fig5} shows how the confidence contours
in the $\Omega_m-h$ and $\Omega_b-h$ planes are improved by including
polarization. The $95\%$ confidence contour  
corresponds roughly to the $68\%$ confidence contour that could be
obtained from temperature 
information alone, while the orientation of the 
error ellipsoids does not change. 
As before the well determined combination 
is  constrained from the temperature data alone.
The constraints on tensor parameters also improve
when polarization is included. Again, this results from 
the better sensitivity to the ionization history, which is 
partially degenerate with the tensor contribution, as discussed 
in the previous section.  
The $B$ channel to which only gravity waves contribute
is not providing additional information in the model
with $T/S=0.28$ for MAP noise levels. Even in a model
with $T/S=1$ the $B$ channel does not provide additional 
information in the case of MAP.

With its very sensitive bolometers, Planck
has the potential to detect the $B$ channel
polarization produced by tensor modes: the 
$B$ channel provides a signature free of scalar mode contribution
(\cite{letter}, \cite{kks}). However, 
it is important to realize that even though for a model
with $T/S=1$ only $20\%$ of the sensitivity of Planck  to tensor
modes is
coming from the $B$ channel. Planck can detected
primordial gravity waves in models 
with $T/S \sim 0.3$ 
in the $B$ channel alone.  If the bolometer sensitivities are improved so that  $w_P^{-1} =
(0.015 \mu K)^2$, then Planck can detect gravity waves
in the $B$ channel even if  $T/S\sim 0.1$. 
We also analyzed the 8-parameter models presented in table 2.
For the models with $T/S=0.28$ and $T/S=0.1$,
MAP will not have sufficient sensitivity to test 
the inflationary consistency relation: $T/S=-7n_T$. Planck
should have sufficient sensitivity to determine $n_T$ with an error 
of 0.2 if $T/S \sim 0.1$, which would  allow a reasonable test 
of the consistency relation. 

Polarization is helping to constrain most of the 
parameters mainly by better constraining  $\tau_{ri}$
and thus removing some of its degeneracies with 
other parameters. Planck will be able to determine not only
the total optical depth through the amplitude of the reionization peak but 
also the ionization fraction, $x_e$,
through its position. To investigate this, we assumed that the 
universe reionized 
instantaneously at $z_{ri}$ and that $x_e$ remains constant but different 
from 1 for
$z<z_{ri}$).
The results given in table 2 indicate 
that $x_e$ can be determined with an accuracy of 15\%. This 
together with the optical depth will be an important test of
galaxy formation models which 
at the moment are consistent with wildly different ionization 
histories and cannot be probed otherwise (\cite{haiman}, \cite{gnedin}).
We also investigated the modified Planck design, where both 
polarization states in bolometers are measured.
An improved polarization noise of $w_P^{-1} =
(0.015 \mu K)^2$ for Planck will shrink the error bars presented
in table 2 by an additional $6-20\, \%$. Error bars on
$\Omega_\Lambda$ and $\Omega_bh^2$ are reduced by $20\%$, those
in $h$, $\tau_{ri}$ and $x_e$  $10-15\, \%$ and for $T/S$, $n_T$
and $n_s$ the improvement is  approximately $6\%$.

\begin{figure}[t]
\centerline{\psfig{figure=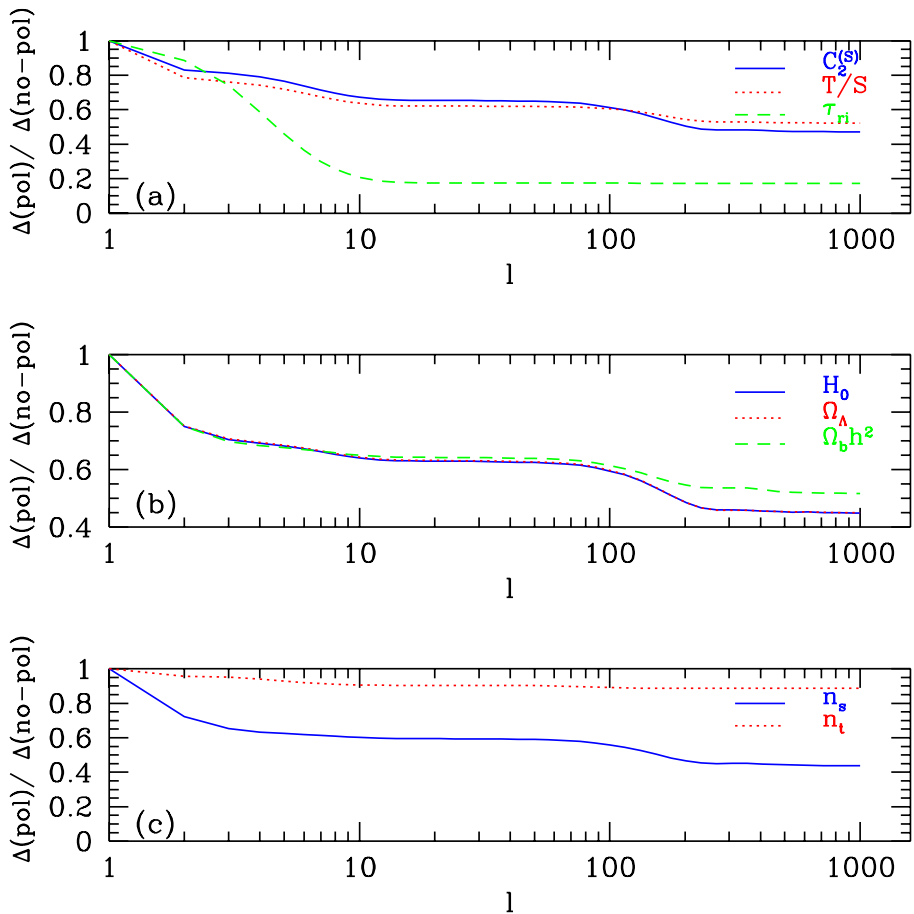,height=5in}}
\caption{Relative improvement in the parameter estimation 
as a function of the maximum $l$ up to which polarization information is
used for the MAP mission.}
\label{fig6} 
\end{figure}

\begin{figure}[t]
\centerline{\psfig{figure=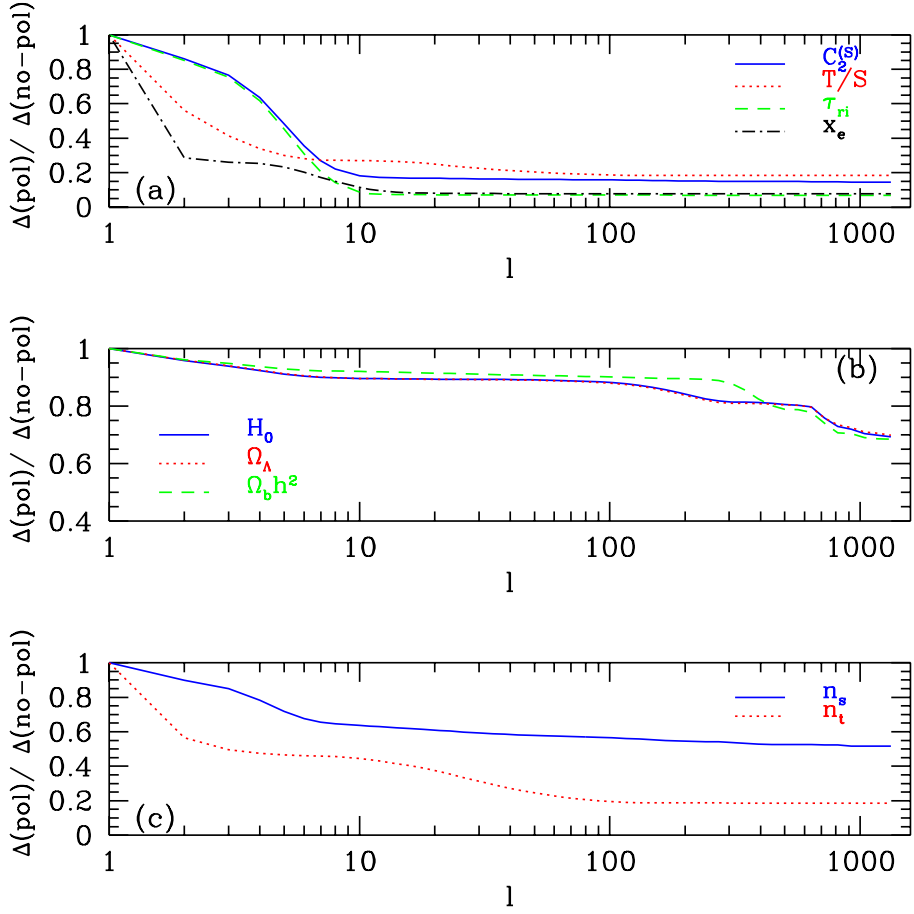,height=5in}}
\caption{Same as figure 6 but for Planck mission 
parameters.}
\label{fig7} 
\end{figure}

We can examine in more detail how polarization helps to constrain 
different cosmological parameters by investigating the angular scales 
in the polarization power spectra that contribute the 
most information. To do
so, we will consider the $T/S=0.1$ model and
perform a Fisher matrix analysis that includes
all the temperature information,  
but polarization information only up to  maximum $l$. 
Figures \ref{fig6} (MAP) and \ref{fig7} (Planck)
show the increase in accuracy as a function of maximum $l$
for various parameters. 
In the case of Planck, we added the ionization fraction after reionization 
as another  parameter.
Most of the increase in information is coming from the low $l$ portion 
of polarization spectrum, primarily from
the peak produced by reionization around $l\sim 10$. 
The first Doppler peak in 
the polarization
spectra at $l\sim 100$ explains the second increase in information
in the MAP case. The better noise properties and resolution 
of Planck  
help to reach the higher $l$ polarization
Doppler peaks, which
add additional information for constraining
$h$, $\Omega_bh^2$ and $\Omega_\Lambda$. For Planck, on the other hand,
some of the degeneracies will already be lifted
in the temperature data alone
and so less is gained when polarization data is used
to constrain the ionization history. 

An interesting question that we can address with 
the methods developed here
is to what extent is one willing 
to sacrifice the sensitivity in temperature to gain sensitivity 
in polarization. A specific example is the 140 GHz channel in Planck, 
where the current proposal is to have
four bolometers with no polarization sensitivity and eight
bolometers which are polarization sensitive so that they transmit only one
polarization state while the other is being thrown away. One 
can compare the
results of the Fisher information matrix analysis for this case 
with the one where
all twelve detectors have only temperature sensitivity, but 
with better overall
noise because no photons are being thrown away. The results 
in this case for
the 8 parameter model with $T/S=0.1$ are 10-20\% better 
than the results given in the fifth column of 
table 1. These results should be compared to the 
same case with polarization in table 2. The latter case is clearly
better for all the parameters, specially for those that are 
degenerate with 
reionization parameters, where the improvement can be quite dramatic.  
Based on this example it seems clear that it is worth 
including polarization sensitivity in the 
bolometer detectors, even at the expense of some sensitivity
in the temperature. However, it remains to be seen whether 
such small levels of polarization can be separated from the 
foregrounds.

\begin{figure}[t]
\centerline{\psfig{figure=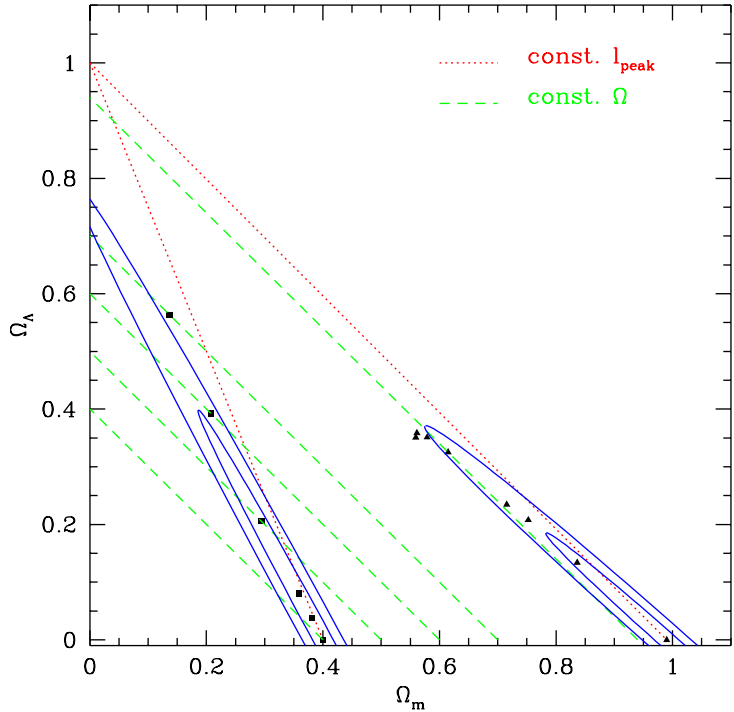,height=4in}}
\caption{Confidence contours $(68\%\  \& \ 95\%)$ in the
$\Omega_\Lambda-\Omega_m$ plane 
for open models in the seven parameter space described in the text.
The dots show the positions of the $\chi^2$ minima found by the
minimization routine when constrained to a subspace of 
constant $\Omega$.}
\label{fig9}
\end{figure}

The Fisher matrix results for the six parameter open models 
are presented in table 2. 
As expected polarization improved the constraints on 
$C_{2}^{(S)}$ and $\tau_{ri}$ the most. So far we have explicitly 
left $\Omega_\Lambda$ out of the analysis; 
as discussed in \S 3 the 
positions of the peaks depends on both 
$\Omega$ and $\Omega_\Lambda$
and it is possible to change 
the two parameters without changing  
the spectrum.
For any given value of $\Omega_m$ we may adjust 
$h$ and $\Omega_b$ to keep  $\Omega_bh^2$ and $\Omega_m h^2$
constant, so that acoustic oscillations will not change. 
If we then in addition adjust also $\Omega_\Lambda$ to match the 
angular size of the acoustic features, then 
the power spectra for two models with different underlying 
parameters remain almost unchanged. 
As mentioned in previous section the effect of 
$\Omega_\Lambda$ on the positions of the peaks 
is rather weak around $\Omega_m=1$ and the peak positions
are mostly sensitive to the curvature $\Omega$. 
The lines of constant $l_{peak}$, the inverse of the angular 
size of acoustic horizon, roughly coincide with those of constant 
$\Omega$ near flat  
models, making it possible to weigh the universe using the
position of the peaks. In the more general case, it is not 
$\Omega$ that can be determined from the CMB observations.
but a particular combination of $\Omega_m$ and $\Omega_\Lambda$
that leaves $l_{peak}$ unchanged.
Figure \ref{fig9} shows  confidence contours in the 
$\Omega_m-\Omega_\Lambda$ plane. The 
contours approximately agree with the constant $l_{peak}$ (dotted) line,
which around $\Omega_m=1$ coincides with the 
constant $\Omega$  line (dashed) 
but not around $\Omega_m=0.4$. 
The squares and triangles correspond to the minima found by the
minimization routine when constrained to move in subspaces of 
constant $\Omega$ and agree with the ellipsoids
from the Fisher matrix approach.   

\begin{figure}[t]
\centerline{\psfig{figure=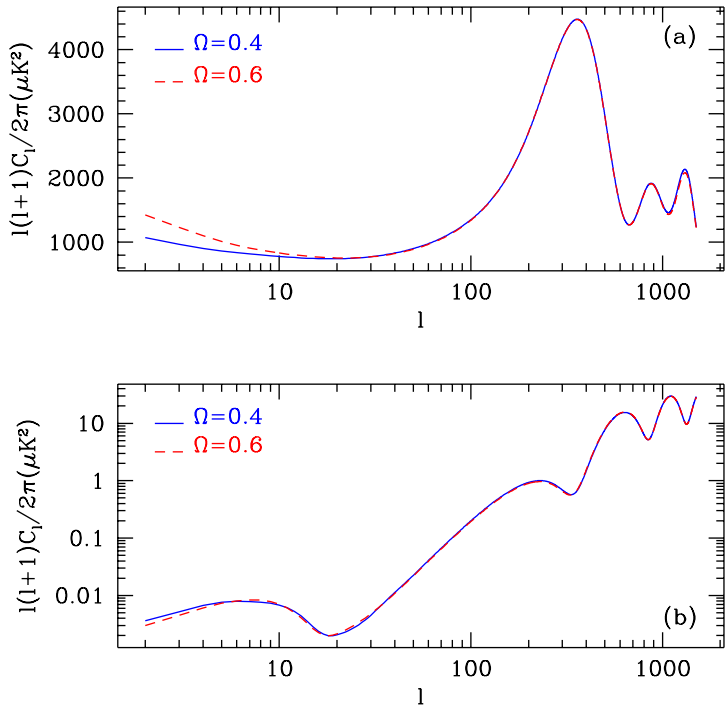,height=4in}}
\caption{Power spectra for 
(a) temperature and 
(b) polarization. The model with $\Omega = 0.6, \Omega_\Lambda=0.4$ is
the output of the minimization code when made to fit the
$\Omega=0.4, \Omega_\Lambda = 0$ model. 
Temperature and polarization data were used for this fit. 
the two models differ in $\chi^2$ by 2.} 
\label{fig10}
\end{figure}

Figure \ref{fig10} shows the temperature and polarization
spectra for the basis model and one found
by the minimization routine with  
$\bi{s}_o=(1495,.87,0.6,0.033,0.051,1.0,0.39)$ where the last number is
now $\Omega_\Lambda$. This model differs
from the basis model by a $\chi^2=2$ 
and so is practically indistinguishable
from it. Only on 
large angular scales do the two models differ somewhat, but 
cosmic variance prevents  an accurate separation between the two.
In this case, polarization does not help to break the degeneracy.
The agreement on the large angular scales is better for polarization 
than for temperature 
because the former does not have a contribution from  
the ISW effect, which is the only effect that can break this degeneracy.
When both $\Omega$ and $\Omega_\Lambda$ are included in the analysis
the $1\sigma$ error bars for both MAP and Planck increase. The greatest
change is for the error bars on the 
curvature that now becomes $\Delta\Omega=0.1$ for
both MAP and Planck.
Note that improving 
the angular resolution does not help to break the degeneracy, which is
why MAP and Planck results are similar. 
If one is willing to allow for both cosmological 
constant and curvature then there is a genuine degeneracy present in
the microwave data and constraints from other cosmological probes 
will be needed to break this degeneracy. 

\section{Shape of the likelihood function, priors and gravitational lensing}

As mentioned in \S 2, the Fisher information matrix approach used so
far assumes that the likelihood function is gaussian around the maximum.
In previous work (\cite{jungman}), this assumption was tested
by calculating the likelihood along several directions in parameter space.
This approach could miss potential problems in other directions, particularly
when there are degeneracies between parameters.
We will further test the gaussian  assumption by 
investigating the shape of the likelihood function varying 
one parameter at a time but marginalizing over the others. 
We fix the relevant parameter and  
let the minimization routine vary all the others in its search for the 
smallest $\chi^2$.  We then repeat the procedure for a different value
of this parameter, mapping the shape of the likelihood function
around the minimum.
The minimization routine is  exploring  parameter space
in all but one direction. These results may be 
compared with the prediction of the
Fisher matrix which follow a parabola in the parameter versus 
log-likelihood plot. 
This comparison tests the gaussianity of the likelihood function
in one direction of parameter space.

\begin{figure}[t]
\centerline{\psfig{figure=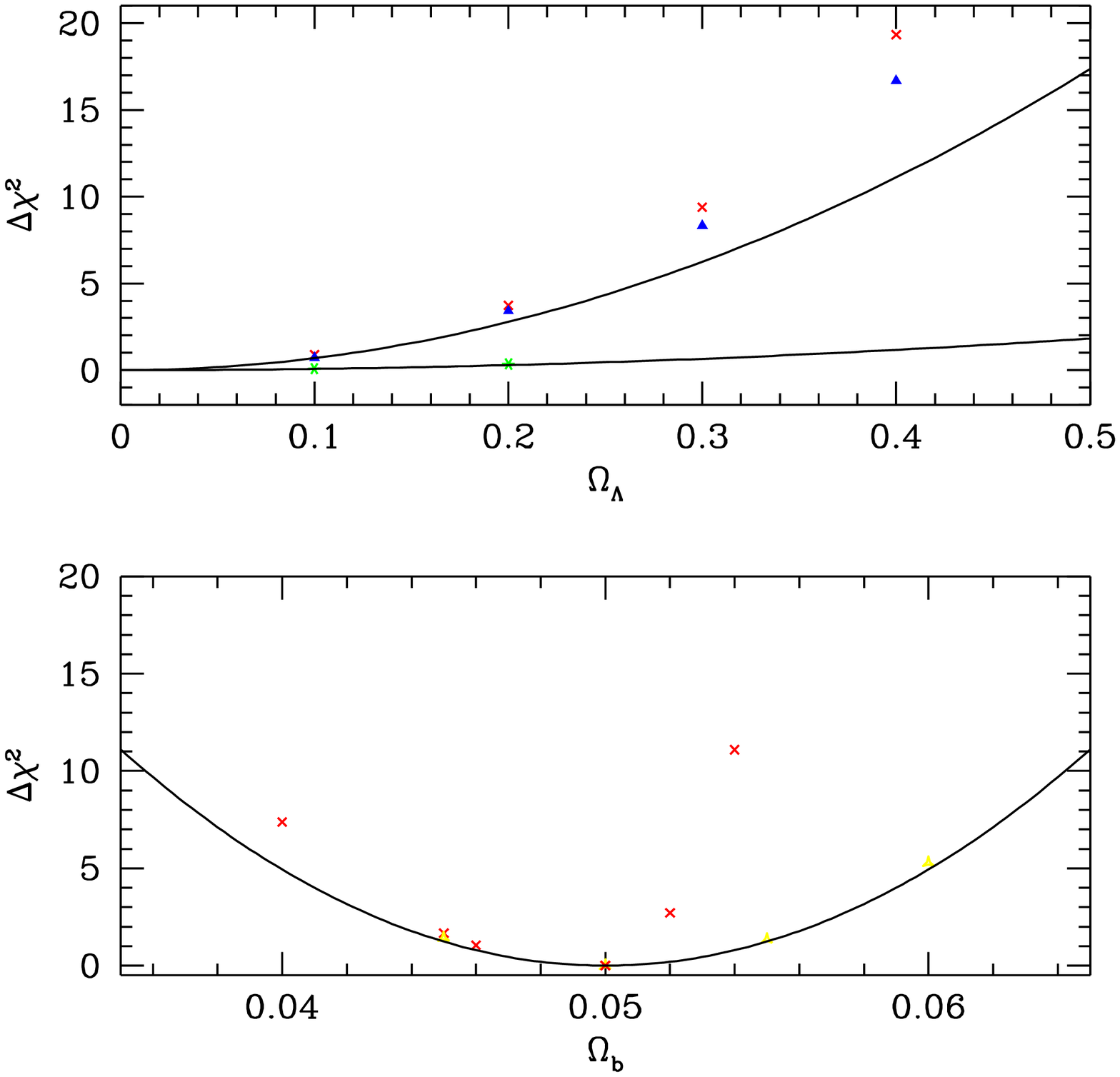,height=5in}}
\caption{Comparison between the Fisher matrix expansion of the likelihood
around the minimum (solid lines) and direct minimization for two
different cosmological parameters. In most cases the agreement near 
the minimum is good. In the upper panel 
full triangles (crosses) 
correspond to fits of sCDM within the six parameter 
family described in the text, including (not including) the effects of
gravitational lensing. The lower curve belongs to the $T/S=0.28$
model in the eight parameter space.
In the lower panel, the $\Omega_\Lambda>0$ prior is reached for 
sCDM when $\Omega_b >0.05$, which is why the minimization results 
differ from the Fisher matrix results.} 
\label{fig12}
\end{figure}

The panels in figure \ref{fig12} show two examples of the results 
of this procedure.
In most cases, the agreement between the
Fisher matrix results and those of the minimization code is very good, 
especially very near the minimum (i.e., $\Delta \chi^2 \lsim 2$). 
As illustrated in the $\Omega_b$ panel, there are cases when 
$\chi^2$ increased more rapidly than predicted by the Fisher matrix. 
This 
is caused by the requirements that $\tau_{ri}$, $\Omega$, $\Omega_\Lambda$
and $\Omega_b$ are all positive, which 
can be enforced easily in the minimization code. 
Of course, such priors are most relevant if the underlying model is
very close to the boundary enforced by the prior and are only important 
on one side of the parameter space.
The importance of this effect therefore depends on the underlying
model. If the amount of information in the CMB data on a given  
parameter is sufficiently high,
then the prior will have only a small effect 
near the maximum of the likelihood function. 

We also investigated the effect of gravitational
lensing on the parameter reconstruction. As shown in Seljak (1996a),
gravitational lensing smears somewhat 
the acoustic oscillations but leaves
the overall shape of the power spectrum unchanged. The amplitude of
the effect depends on the power spectrum of density fluctuations. 
Because the CMBFAST output consists of both CMB and density power spectra
one can use them as an input for the 
calculation of the weak lensing effect following the method in 
Seljak (1996a). The gravitational lensing effect is treated 
self-consistently by normalizing the power spectrum for each model 
to COBE. 
We find that the addition of gravitational
lensing to the calculation does not appreciably 
change the expected sensitivity 
to different parameters that will be attained
with the future CMB experiments. This conclusion again depends somewhat
on the underlying model, but even for sCDM where  COBE
normalization predicts two times larger small scale normalization 
than required by the cluster abundance data, the lensing effect 
is barely noticeable in the error contours for various parameters.
 
\section{Conclusions}

In this paper,
we have analyzed how accurately cosmological parameters can
be extracted from the CMB measurements
by two future satellite missions. Our work differs from previous
studies on this subject  in that we
use a more accurate 
computational code for calculating the theoretical spectra and we 
include the additional information that is present in the polarization 
of the microwave background. We also investigate how the results 
change if we vary the number of parameters to be modeled 
or the underlying model around which the parameters are 
estimated. Both of these variations can have a large effect 
on the claimed accuracies of certain parameters, so the numbers presented
here should not be used as firm numbers but more as
typical values. Of course, once the underlying model is revealed
to us by the observations then
these estimates can be made more accurate.
The issue of variation of the errors on the number of parameters however 
remains, and  results will always 
depend to some extent on the prior belief. If, for example,
one believes that gravity waves are not generated in inflationary
models (e.g., \cite{lyth}) or that they are related to the scalar 
perturbations through a simple relation (e.g., \cite{turner}),
then the MAP errors on most parameters shrink by
a factor of 2. Similarly, one may decide that models with 
both curvature and cosmological constant are not likely, which removes
the only inherent degeneracy present in the CMB data.

Using temperature data alone, MAP should be able to make
accurate determinations (better than 10\%) of the scalar
amplitude ($\sigma_8$), the baryon/photon ratio
($\Omega_b h^2$), the matter content ($\Omega_m h^2)$,
the power spectrum slope ($n_s$) and the angular
diameter distance to the surface of last scattering (a combination
of $\Omega$ and $\Omega_\Lambda$).  If we restrict ourselves
to models with no gravity wave content, then MAP should also be able
to make accurate determinations of $\Omega_\Lambda$,
the Hubble constant and the optical depth, $\tau_{ri}$.  However,
in more general models that include gravity wave amplitude and
slope as additional parameters, the degeneracies between these parameters
are large and they cannot be accurately determined. Several other 
measurements of the CMB anisotropies from the ground and from 
balloons are now in progress and very accurate results are likely to
be available by the time MAP flies. This additional information 
will help constrain the models further, specially determinations of 
the power spectrum at the smaller angular scales.   
Astronomical data can significantly reduce these degeneracies.
The two nearly degenerate models, sCDM and a tilted
vacuum dominated model ($\Omega_\Lambda = 0.6$) shown in figure 2
can be distinguished already by current 
determinations of $\sigma_8 \Omega^{0.6}$, or by measurements of
the shape parameter $\Gamma$ in the galaxy power spectrum,
or by measurements of the distance-magnitude relationship with
SN Ia's.

MAP's measurements of polarization will significantly enhance
its scientific return.  These measurements will accurately 
determine the optical depth between the present and the surface of last
scatter.  This will not only probe star formation during the
``dark ages'' ($5 < z < 1300$), but will also enable accurate determination
of the Hubble constant and help place interesting constraints
on $\Omega_\Lambda$ in models with tensor slope and amplitude
as free parameters.

Planck's higher sensitivity and smaller angular resolution
will enable further improvements in the parameter determination.
Particularly noteworthy is its ability to constrain $\Omega_\Lambda$
to better than 5\%  and the Hubble constant
to better than 1\% even in the most general
model considered here.  
The proposed addition of polarization sensitive bolometer channels
to  Planck significantly enhances its science return. Planck 
should be able to measure the ionization
history of the early universe, thus studying primordial star formation.
Sensitive polarization measurements should enable Planck to determine
the amplitude and slope of the gravity wave spectrum. This is particularly
exciting as it directly tests
the predicted tensor/scalar relations in the inflationary theory
and is a probe of Planck scale physics.  
The primordial gravity wave contribution can at present only be measured
through the CMB observations. One may therefore ask  how
well Planck can determine $T/S$ assuming that other cosmological parameters 
are perfectly known by combining CMB and other astronomical data. The
answer sensitively depends on reionization optical depth. Without 
reionization, $T/S \sim 0.1$ can be detected, while with $\tau_{ri}=0.1$
this number drops down to $T/S \sim 0.02$. The equivalent number without 
polarization information is $0.2$,
regardless of optical depth.
Improvements in 
sensitivity will further improve these numbers, particularly 
in the $B$ polarization channel which is not cosmic variance
limited in the sense that tensors cannot be confused with scalars.
A detection of a $B$ component
would mean a model independent detection of a stochastic
background of gravitational waves or vector modes
(Seljak \& Zaldarriaga 1997).

The most exciting science return from polarization measurements
come  from measurements at large angular scales (see figures 6 \& 7).  
These measurements can only be made
from satellites as systematic effects will swamp balloons
and ground based experiments on these scales.  The low $l$ measurements
enable determinations of the optical depth and the ionization history
of the universe and may lead to the detection of gravity
waves from the early universe. Both foregrounds and systematic
effects may swamp the weak polarization signal, even in
space missions, thus it is important that the satellite experiment
teams adopt scan strategies and frequency coverages 
that can minimize systematics and foregrounds at large
angles.

We explored the question of how priors such as positivity of 
certain parameters or constraints
from other cosmological probes
help reduce the uncertainties from the CMB data alone.
For this purpose, we compared the predictions from the Fisher information 
matrix  with those of the brute-force minimization which allows
the  easy incorporation of inequality priors. As expected, 
we find that positivity changes the 
error estimates only on the parameters that are not well constrained 
by the CMB data. On the other hand, using some additional constraints
such as the limits on the Hubble constant, age of the universe, 
dark matter power spectrum or $q_0$ measurements from 
type Ia supernovae can significantly reduce the error estimates
because the degeneracies present in these cosmological tests are typically
different from those present in the CMB data.
The minimization approach also allows testing the 
assumption that the log-likelihood is well described by a quadratic 
around the minimum, which is implicit 
in the Fisher matrix approach.
We find that this
 is a good approximation close to the minimum, with no 
nearby secondary minima that could be confused with the global one. 
Finally, we also tested the effect of gravitational lensing on the 
reconstruction of parameters and found that its effect on the 
shape of the likelihood function can be neglected. 

In summary, future CMB data will provide us with an unprecedented
amount of information in the form of temperature and polarization 
power spectra. Provided that the true cosmological model belongs
to the class of models studied here these data will enable us to constrain 
several combinations of cosmological parameters with an exquisite
accuracy. While some degeneracies between the cosmological
parameters do exist, and in principle do not allow some of them to be 
accurately determined individually, these can be removed by including 
other cosmological constraints. Some of 
these degeneracies
belong to contrived cosmological models, which may not survive 
when other considerations are included. 
The microwave background is at present
our best hope for an accurate determination of classical cosmological 
parameters.

\acknowledgements
We would like to thank C. Bennett, M. Kamionkowski and M. Tegmark
for helpful comments. M.Z. would like to thank the hospitality of 
the Institute for Advance Study
where the final part of this work was performed.

\newpage
\oddsidemargin -0.5in
\begin{tabular}{||l|l|l|l|l|l|l|l|l||} \hline
$Param.$&${\rm sCDM}^+$&${\rm sCDM}^+$&${T \over S}=0.28^+$&${T \over S}=0.1^+$ 
&${T \over S}=0.1^\times$&$\Omega_\Lambda=0.7^+$
& Open$^+$ & Open$^\times$\\ \hline
$\Delta \ln C_2^{(S)}$ & $2.1\, 10^{-1}$ & $4.2\, 10^{-1}$& 
$4.8\, 10^{-1}$ 
& $4.7 \, 10^{-1}$ & $7.4\, 10^{-2}$&$ 4.1\, 10^{-1}$
& $1.2 \, 10^{-1}$ & $4.7 \, 10^{-2} $ \\ \hline
$\Delta h$& $1.7\, 10^{-2}$  & $9.2\, 10^{-2}$& 
$1.1\, 10^{-1}$ & $1.0 \, 10^{-1}$ 
& $5.1\, 10^{-3}$ &$ 4.1\, 10^{-2}$ & $2.0 \, 10^{-2}$ & 
$1.1 \, 10^{-3}$ \\ \hline
$\Delta \Omega_\Lambda$& $9.8 \, 10^{-2}$& $5.3 \, 10^{-1}$ 
& $6.1 \, 10^{-1}$ & $5.8 \, 10^{-1}$& $2.9 \, 10^{-2}$ 
& $5.0 \, 10^{-2}$ & - & -\\ \hline
$\Delta \Omega_b h^2$& $3.0\, 10^{-4}$ & $1.0\, 10^{-3}$ & 
$9.8\, 10^{-4}$ & $1.2\, 10^{-3}$ 
& $1.2\, 10^{-4}$ & $9.7\, 10^{-4}$
& $1.1 \, 10^{-3}$ & $1.3 \, 10^{-4}$ \\ \hline
$\Delta \tau_{ri}$& $1.2\, 10^{-1}$ & $1.3\, 10^{-1}$ & 
$1.4\, 10^{-1}$& $1.1\, 10^{-1}$&$8.2 \, 10^{-2}$ &$1.9 \, 10^{-1}$
& $7.2 \, 10^{-2}$ & $3.3 \, 10^{-2}$ \\ \hline
$\Delta n_s$& $9.8\, 10^{-3}$ & $5.9\, 10^{-2}$ & 
$6.7\, 10^{-2}$ & $6.4\, 10^{-2}$ & $5.9\, 10^{-3}$ & $2.9\, 10^{-2}$
& $2.4\, 10^{-2}$ & $3.3 \, 10^{-3}$\\ \hline
$\Delta {T \over S}$& - & $3.9\, 10^{-1}$ & $6.8\, 10^{-1}$ & 
$5.3 \, 10^{-1}$ & $2.5 \, 10^{-1}$ & $3.2 \, 10^{-1}$& - & -\\ \hline
$\Delta n_T$& - & - & $3.9\, 10^{-1}$ & $9.1\, 10^{-1}$ &
$9.4\, 10^{-1}$ & $9.9 \, 10^{-1}$& - & -\\ \hline
$\Delta \Omega$& - & - & - 
&- & - &- & $6.6\, 10^{-3}$& $5.2\, 10^{-4}$ \\ \hline
\end{tabular}

\vspace{.5cm}
\noindent{Table 1. Fisher matrix one-sigma error bars for different
cosmological parameters when only temperature is included. 
Table 3 gives the cosmological parameters for each of the models.
Columns with $+$ correspond to MAP and those
with $\times$ to Planck.} 

\vspace{1cm}
\begin{tabular}{||l|l|l|l|l|l|l|l|l||} \hline
$Param.$&${\rm sCDM}^+$&${\rm sCDM}^+$&${T \over S}=0.28^+$&${T \over S}=0.1^+$ 
&${T \over S}=0.1^\times$&$\Omega_\Lambda=0.7^+$
& Open$^+$ & Open$^\times$\\ \hline
$\Delta \ln C_2^{(S)}$ & $4.8\, 10^{-2}$ & $2.4\, 10^{-1}$& 
$2.8\, 10^{-1}$ & $2.4 \, 10^{-1}$ & $1.0\, 10^{-2}$& $8.3\, 10^{-2}$
&$6.5\, 10^{-2}$ & $1.2\, 10^{-2}$\\ \hline
$\Delta h$& $1.6\, 10^{-2}$  & $5.1\, 10^{-2}$& 
$5.8\, 10^{-2}$ & $5.0\, 10^{-2} $ & $3.0\, 10^{-3}$& $3.8\, 10^{-2}$
&$1.9\, 10^{-2}$ & 
$1.0\, 10^{-3}$\\ \hline
$\Delta \Omega_\Lambda$& $9.3 \, 10^{-2}$& $2.9 \, 10^{-1}$ 
& $3.3 \, 10^{-1}$ & $2.9 \, 10^{-1}$& $1.7 \, 10^{-2}$ & $4.6\, 10^{-2}$
& - & - \\ \hline
$\Delta \Omega_b h^2$
& $2.8\, 10^{-4}$ & $6.1\, 10^{-4}$ & 
$7.1\, 10^{-4}$ & $6.2\, 10^{-4}\ $ & $5.7\, 10^{-5}\ $ & $8.9\, 10^{-4}$&
$9.5\, 10^{-4}$ & $1.1\, 10^{-4}$\\ \hline
$\Delta \tau_{ri}$& $2.1\, 10^{-2}$ & $2.1\, 10^{-2}$ & 
$2.0\, 10^{-2}$& $2.0\, 10^{-2}$&$5.5 \, 10^{-3}$
& $2.0\, 10^{-2}$&$3.2 \, 10^{-2}$ & $ 3.5\, 10^{-3}$\\ \hline
$\Delta n_s$& $4.8\, 10^{-3}$ & $3.1\, 10^{-2}$ & 
$3.5\, 10^{-1}$ & $3.0\, 10^{-2}$ & $3.0\, 10^{-3}$
& $2.6\, 10^{-2}$&$1.7 \, 10^{-2}$ & $2.6\,  10^{-3}$\\ \hline
$\Delta {T \over S}$& - & $2.2\, 10^{-1}$ & $4.3\, 10^{-1}
$ & 
$3.0 \, 10^{-1}$ & $4.5 \, 10^{-2}$& $2.1\, 10^{-1}$& - & -\\ \hline
$\Delta n_T$& - & - & $3.9\, 10^{-1}$ & $8.1\, 10^{-1}$ &
$1.7\, 10^{-1}$ & $7.8\, 10^{-1}$& - & -\\ \hline
$\Delta x_e$& - & - & -& - & $1.4\, 10^{-1}$& - & - &-\\ \hline
$\Delta \Omega$& - & - & - 
&- & - &- & $6.1\, 10^{-3}$& $4.1\, 10^{-4}$ \\ \hline
\end{tabular}

\vspace{0.5cm}
\noindent{Table 2. Fisher matrix one-sigma error bars for different
cosmological parameters when both temperature and polarization is included. 
Table 3 gives the cosmological parameters for each of the models.
Columns with $+$ correspond to MAP and those
with $\times$ to Planck.}

\vspace{1cm}
\begin{tabular}{||l|l|l|l|l|l||} \hline
$Param.$&${\rm sCDM}$&${T \over S}=0.28$&${T \over S}=0.1$ 
&$\Omega_\Lambda=0.7$
& Open \\ \hline
$h$& $0.5$  & $0.5$& 
$0.5$ & $.65$ & $.65$\\ \hline
$\Omega_\Lambda$& $0.0$& $0.0$ 
& $0.0$ & $0.7$& $0.0$ \\ \hline
$\Omega_b $
& $0.05$ & $0.05$ & 
$0.05$ & $0.06$ & $0.06$\\ \hline
$\tau_{ri}$& $0.05$ & 
$0.05$& $0.1$&$0.1$
& $0.05$\\ \hline
$n_s$& $1.0$ & $0.96$ & 
$0.99$ & $1.0$ & $1.0$\\ \hline
${T \over S}$& $0.0$ & $0.28$& $0.1$ & 
$0.0$ & $0.0$\\ \hline
$n_T$& - & $0.04$ & $0.01$ & - & -\\ \hline
$\Omega$& $1.0$ & $1.0$ & $1.0$ & $1.0$ & $0.4$ \\ \hline
\end{tabular}

\vspace{0.5cm}
\noindent{Table 3. Cosmological parameters for the models we studied.
All models were normalized to COBE.}

\end{document}